\documentclass[preprints,review,accept,moreauthors,pdftex,10pt,a4paper]{Definitions/mdpi}
\firstpage{1}
\pubvolume{xx}
\issuenum{1}
\articlenumber{5}
\pubyear{2019}
\copyrightyear{2019}
\history{Received: 27 November 2018; Accepted: 23 December 2018; Published: date}
\updates{yes} % If there is an update available, un-comment this line

\usepackage{aas_macros}
\usepackage{amssymb}
\usepackage{xspace}

\graphicspath{{./},{./figs/}}

\newcommand{\flaapluc}{\texttt{FLaapLUC}\xspace}
\newcommand{\hess}{H.E.S.S.\xspace}
\newcommand{\lat}{\textit{Fermi}-LAT\xspace}
\newcommand{\g}{\ensuremath{\gamma}\xspace}

\Title{Monitoring the Extragalactic High Energy Sky}

\Author{Jean-Philippe Lenain\orcidA{} for the \hess Collaboration}
\AuthorNames{Jean-Philippe Lenain for the \hess collaboration}

\address[1]{%
Sorbonne Université, Université Paris Diderot, Sorbonne Paris Cité, CNRS/IN2P3, Laboratoire de Physique Nucléaire et de Hautes Energies, LPNHE, 4 Place Jussieu, F-75252 Paris, France; contact.hess@hess-experiment.eu}

\abstract{Blazars are jetted active galactic nuclei with a jet pointing close to the line of sight, hence~enhancing their intrinsic luminosity and variability. Monitoring these sources is essential in order to catch them flaring and promptly organize follow-up multi-wavelength observations, which are key to providing rich data sets used to derive e.g.,  the emission mechanisms at work, and the size and location of the flaring zone. In this context, the \lat has proven to be an invaluable instrument, whose data are used to trigger many follow-up observations at high and very high energies. A few examples are illustrated here, as well as a description of different data products and pipelines, with a focus given on \flaapluc, a tool in use within the \hess collaboration.
}

\begin{document}

%%%%%%%%%%%%%%%%%%%%%%%%%%%%%%%%%%%%%%%%%%
\section{Introduction}

Active galactic nuclei (AGN) are among the most energetic sources in the Universe. They~harbor a supermassive black hole in the center of their host galaxy, fueled by an accretion disk emitting mainly UV and X-ray thermal radiation. Blazars are a specific class of AGN, exhibiting relativistic jets, with~emission dominated by non-thermal radiation from the radio to the \g-ray bands. The~observational peculiarity of blazars stems from the quasi-alignment of their jet with our line of sight, thus Doppler boosting their emission \citep{1966Natur.211..468R,1978PhyS...17..265B}. Thus, their luminosity can be highly variable in almost all wavebands, and this variability can occur on a broad range of time scales, from years down to minutes.

Blazars are further divided in sub-classes:~flat spectrum radio quasars (FSRQ), low-frequency-peaked BL\,Lac objects (LBL), intermediate-frequency-peaked BL\,Lac objects (IBL) and high-frequency-peaked BL\,Lac objects (HBL), depending on the peak frequency of their synchrotron component, but which also display different luminosities. FSRQ exhibit emission lines in the optical/UV range from e.g.,  their accretion disk, while BL\,Lac objects are mostly devoid of any detectable thermal component. From these different types, Fossati et al. \cite{1998MNRAS.299..433F} proposed the existence of a blazar sequence, characterized by an anti-correlation between peak frequency and luminosity. Since~then, several studies have been conducted to assess whether this sequence is real or due to selection effects (see, e.g., \citep{2007Ap+SS.309...63P,2008A+A...488..867N,2008MNRAS.387.1669G,2008MNRAS.391.1981M,2011ApJ...740...98M,2012MNRAS.420.2899G}). In the meantime, several outliers from this sequence have been detected, with, for instance, objects exhibiting both a high peak frequency and bolometric luminosity (e.g., \citep{2012MNRAS.422L..48P,2017A+A...606A..68C}). This subject is still a matter of debate (see, e.g., \citep{2017MNRAS.469..255G,2017A+ARv..25....2P}, and references therein).

\begin{figure}[H]
\centering
\includegraphics[width=0.7\textwidth]{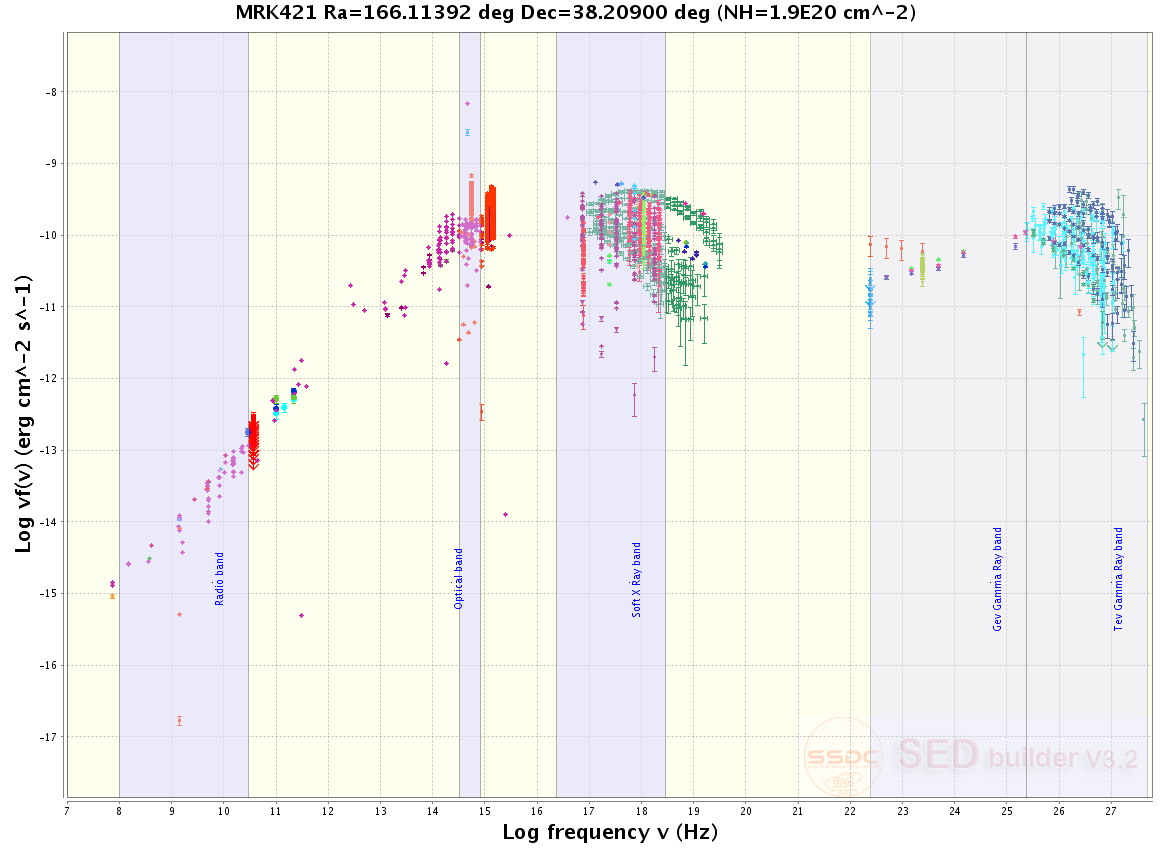}
\caption{Spectral energy distribution of Mrk\,421, showing the two typical spectral components, and~variability, generated using
\href{https://tools.ssdc.asi.it/SED}{https://tools.ssdc.asi.it/SED}, data from \cite{1970ApJS...20....1D,1981MNRAS.197..865W,1992ApJS...79..331W,1992ApJS...80..257E,1996ApJS..103..427G,1997ApJ...475..479W,1998AJ....115.1693C,1999A+A...349..389V,1999ApJS..123...79H,2002babs.conf...63G,2003ApJ...598..242A,2003MNRAS.341....1M,2005A+A...437...95A,2007A+A...472..705V,2007AJ....133.1947N,2007ApJS..171...61H,2007MNRAS.376..371J,2008A+A...480..611S,2009A+A...506.1563P,2010A+A...510A..48C,2010A+A...524A..64C,2010AJ....140.1868W,2010ApJS..186....1B,2010ApJS..188..405A,2010JPhG...37l5201C,2011A+A...536A...7P,2011ApJ...734..110B,2011ApJ...738...25A,2011PASJ...63S.677H,2012ApJ...749...21A,2012ApJS..199...31N,2012JPhG...39d5201C,2013ApJ...779...27B,2013ApJS..207...19B,2013ApJS..207...36H,2014A+A...571A..28P,2014ApJS..210....8E,2015ApJ...812...60B,2015ApJS..218...23A,2015NIMPA.770...42S,2016A+A...588A.103B,2016A+A...590A...1R,2016A+A...594A..26P}.}
\label{fig-mrk421_sed}
\end{figure}

The spectral energy distribution (SED) of blazars typically shows two main components (see, e.g.,  Figure~\ref{fig-mrk421_sed}). A low-energy component, peaking in the infrared to X-ray band, is commonly ascribed to synchrotron emission from relativistic electrons and positrons in the jet. The nature of the high energy component, peaking at high energies, can depend on the source properties and/or activity state. Leptons from the jet can interact with their own synchrotron emission by inverse Compton scattering, via the synchrotron self-Compton mechanism \citep{1965ARA+A...3..297G,1981ApJ...243..700K}. In FRSQ or LBL objects, the thermal emission from the accretion disk, from the broad line region or the dusty torus can contribute significantly and give rise to external inverse Compton emission at high energies. In a hadronic framework, proton~synchrotron and $p$--\g interactions can dominate the high energy emission (see, e.g., \citep{1991A+A...251..723M,2001APh....15..121M}).

The study of the variability in blazars, essentially of their flares, can help to reveal the nature of the emission mechanisms, as well as the location of the source emission (e.g., \citep{2015ApJ...815L..22A,2017ICRC...35..655Z,2017ICRC...35..649R}). By contrast, in order to detect the thermal emission, investigable mostly in the near-infrared/optical range, the~sources should be caught while in a quiet non-thermal state, such that the usually dominating non-thermal component does not outshine the thermal one. This is an almost necessary condition to reliably measure their redshift, obviously an essential quantity to derive the energetic budget in these objects (see, e.g.,~\citep{2013ApJ...764..135S,2017AIPC.1792e0025P}).

\section{Facilities and Coverage}

\begin{figure}[H]
\centering
\includegraphics[width=0.7\textwidth]{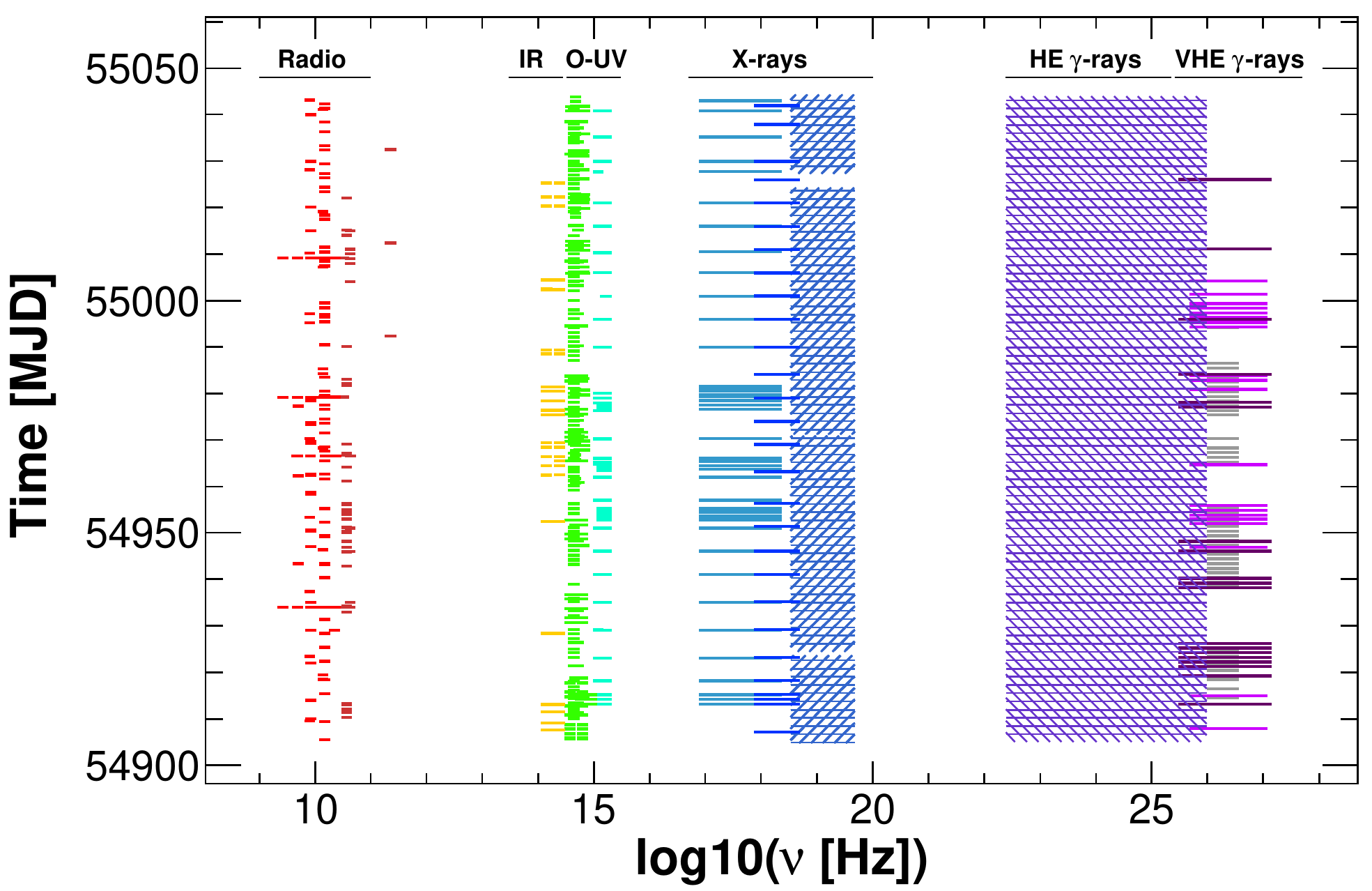}
\caption{Time and energy band coverage of the multi-wavelength campaign on Mrk\,501 in 2009, from~Abdo et al.~\cite{2011ApJ...727..129A}.}
\label{fig-mrk501_mwl_coverage}
\end{figure}

Catching AGN in flaring state, or quieter activity state for redshift measurements for instance, requires monitoring them with high cadence, given the diverse variability time scales they display. However, most astrophysical facilities have small fields of view, with respect to the entire sky, and rare are the instruments capable of instantly observing a sizable fraction of the celestial vault. Furthermore, depending on the observed wavelength and modes of detection, most instruments can not continuously operate. Such low duty cycles and small fields of view make exhaustive monitoring programs difficult to achieve, either by selecting a few sources on which observational efforts are focused, or due to high time pressure on the overall facility's observation program. In a few cases, it is possible to gather different observatories to observe the same source at almost the same time (see, e.g.,  Figure~\ref{fig-mrk501_mwl_coverage}), at the expense of notable efforts. This has been brilliantly demonstrated very recently during the multi-wavelength and multi-messenger campaign on NGC\,4993 at the occasion of the binary neutron star merger event GW\,170817 \cite{2017ApJ...848L..12A}, or for the campaign organized on the presumable electromagnetic counterpart TXS\,0506$+$056 of the neutrino event IceCube-170922A \cite{2018Sci...361.1378I}. However, observations performed with facilities having small fields of view and/or low duty cycle tend to be naturally biased to observations of high-flux states, if no long-term, persevering monitoring programs with regular cadence are in place.

\begin{figure}[H]
\centering
\includegraphics[width=0.86\textwidth]{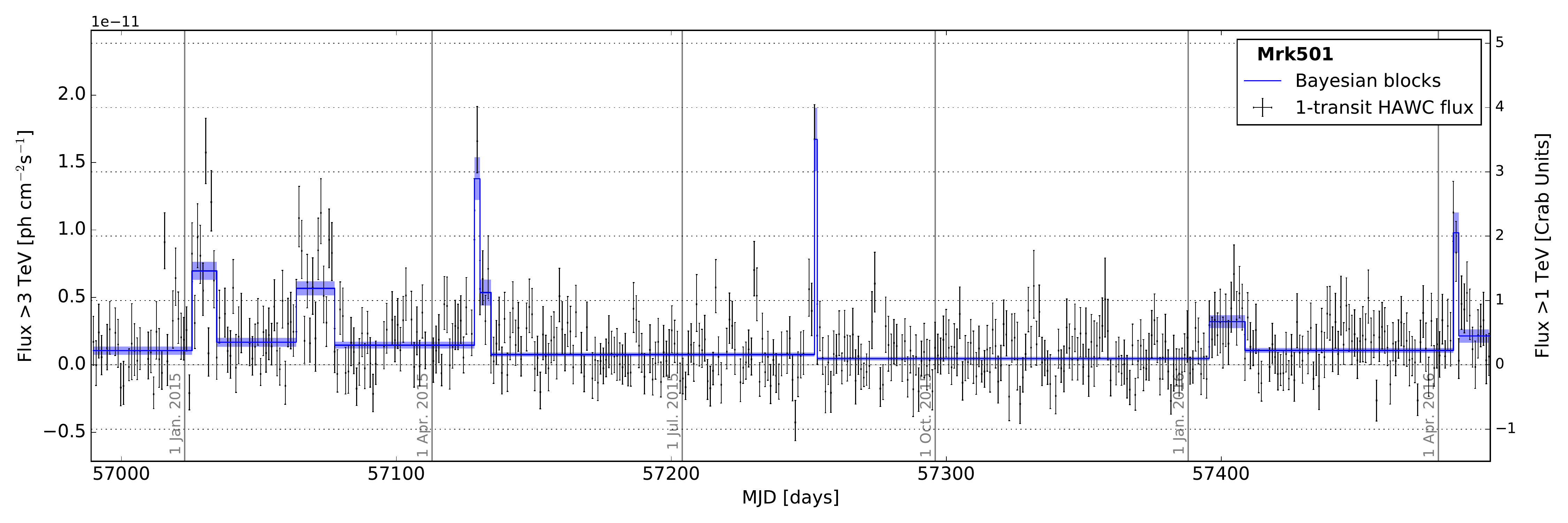}\\
\includegraphics[width=\textwidth]{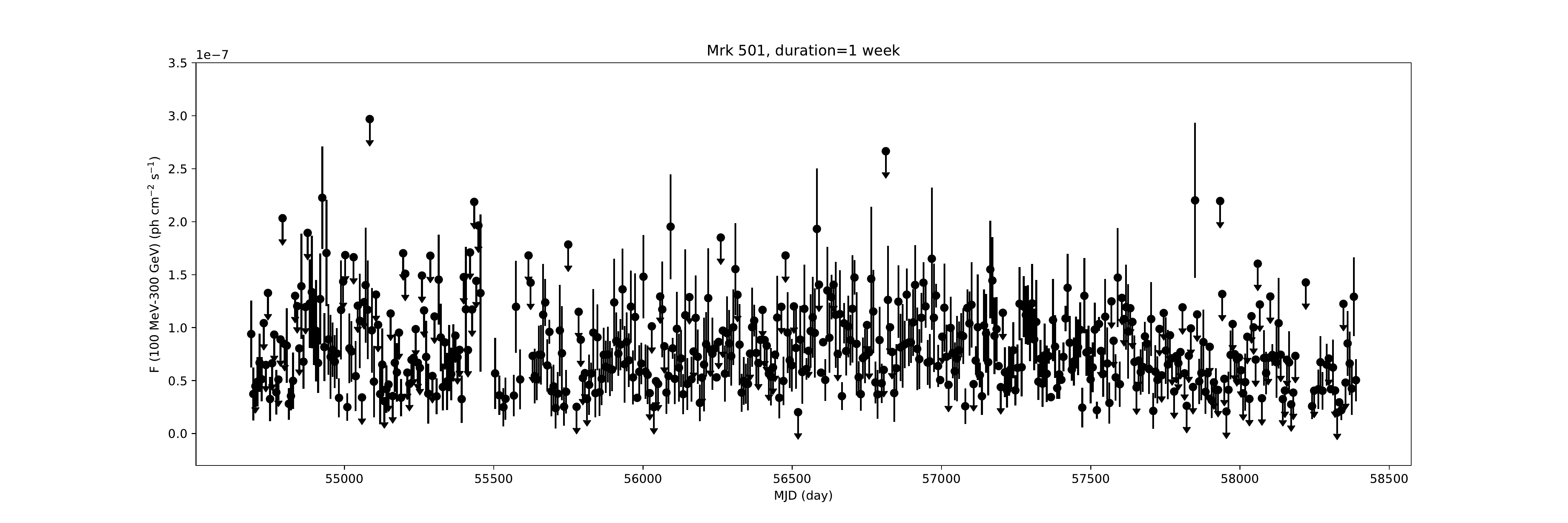}\vspace{-6pt}
\caption{(Top) Light curve of Mrk\,501 as seen with HAWC from November 2014 to April 2016, with~a daily sampling. The blue lines show distinct flux states from a Bayesian block analysis, from~Abeysekara et al.~\cite{2017ApJ...841..100A};
(bottom) long-term, weekly-binned light curve of Mrk\,501, as observed with \lat for 10 years. Adapted from~\cite{msllc}.}
\label{fig-mrk501_lc}
\end{figure}

On the opposite scale, some facilities such as the \textit{Neil Gehrels Swift}/BAT \citep{2005SSRv..120..143B}, \textit{AGILE} \citep{2009A+A...502..995T}, \textit{Fermi}-GBM \citep{2009ApJ...702..791M} and \lat \citep{2009ApJ...697.1071A} or HAWC \citep{2017ApJ...843...39A} have
%Define LAT and HAWC also if appropriate
all-sky capabilities, with a large field of view, and almost continuous or high cadence observations (see Figure~\ref{fig-mrk501_lc}). Without such monitoring capabilities, it would have been, for instance, barely possible to serendipitously discover flares from the Crab Nebula, previously thought to be a steady \g-ray source (see, e.g., \citep{2014RPPh...77f6901B}, and references therein). Furthermore, at high Galactic latitudes, the high energy \g-ray sky is largely dominated by AGN, and~particularly blazars \citep{2015ApJS..218...23A}, making \textit{AGILE} and \lat perfect instruments to monitor the high energy extragalactic sky.

In particular, the \lat has proven to be very useful to alert the community on unusual events for follow-up observations. As examples, observations at very high energies (VHE; $E \gtrsim 100$\,GeV) following up source activity detected at high energies with the \lat yielded the following recent~results:

\begin{itemize}[leftmargin=8mm,labelsep=5.5mm]
    \item the detection of the FSRQ Ton\,599 by both MAGIC \cite{2017ATel11061....1M} and VERITAS \cite{2017ATel11075....1M}, following an alert from \lat \cite{2017ATel10931....1C};
    \item the discovery of VHE emission from a gravitationally lensed blazar, S3\,0218$+$357, at the time of arrival of the lensed component \cite{2014ATel.6349....1M,2016A+A...595A..98A}, which followed the prompt flare reported by~\lat~\cite{2014ATel.6316....1B}, making it the furthest VHE \g-ray source known to date with $z=0.944$;
    \item the detection of the second furthest VHE \g-ray source (with $z=0.939$), the FSRQ PKS\,1441$+$25, with MAGIC \cite{2015ATel.7416....1M} and VERITAS \cite{2015ATel.7433....1M} following activity detected in the optical, X-ray and high-energy \g-ray ranges \cite{2015ATel.7402....1P};
    \item the detection at VHE of the FSRQ PKS\,0736$+$017 with \hess---Ref. \cite{2017AIPC.1792e0029C} following a \g-ray flare identified as explained in the following.
\end{itemize}

%%%%%%%%%%%%%%%%%%%%%%%%%%%%%%%%%%%%%%%%%%
\section{\textit{Fermi}-LAT as an All-Sky Monitor and Flare Detector}

The \lat team and the Fermi Science Support Center (FSSC) at NASA provide numerous data products and analysis tools to the community, from the Fermitools\footnote{\href{https://fermi.gsfc.nasa.gov/ssc/data/analysis/software/}{https://fermi.gsfc.nasa.gov/ssc/data/analysis/software}} needed to analyze
%Pls move the footnotes to the main section of the paper; footnotes are not allowed
data, to~high-level data products\footnote{\href{https://fermi.gsfc.nasa.gov/ssc/data/access/lat/}{https://fermi.gsfc.nasa.gov/ssc/data/access/lat}}. Catalog products, such as those for the 3FGL \cite{2015ApJS..218...23A} or 3FHL \cite{2017ApJS..232...18A}, are~also made available.

Of interest for source monitoring and potential follow-up observations on relevant source activities are the available light curves for every source having experienced a flaring activity with a flux exceeding $10^{-6}$\,cm$^{-2}$\,s$^{-1}$ at least once during the \lat mission lifetime \footnote{\href{https://fermi.gsfc.nasa.gov/ssc/data/access/lat/msl_lc/}{https://fermi.gsfc.nasa.gov/ssc/data/access/lat/msl\_lc}}, as well as the light curves obtained using aperture photometry of all sources belonging to the 3FGL catalog\footnote{\href{https://fermi.gsfc.nasa.gov/ssc/data/access/lat/4yr_catalog/ap_lcs.php}{https://fermi.gsfc.nasa.gov/ssc/data/access/lat/4yr\_catalog/ap\_lcs.php}}. It should be noted, though, that these light curves have no absolute flux calibration but are relevant to reveal changes of activity, i.e., relative flux variations, from a given object. Another interesting product is given by the Fermi All-sky Variability Analysis (FAVA) tool\footnote{\href{https://fermi.gsfc.nasa.gov/ssc/data/access/lat/FAVA/}{https://fermi.gsfc.nasa.gov/ssc/data/access/lat/FAVA}} \cite{2013ApJ...771...57A,2017ApJ...846...34A}, which blindly searches the whole sky to detect flares from \g-ray sources. One caveat related to the use of this tool is that its latency for flare detection is of the order of a week in order to accumulate enough statistics, which is sufficient for long-lasting events, but too long for e.g.,  some AGN flares which may last only a few days or less.

As a result, one may be interested in piping such products into an automatic stream such that follow-up observations could be promptly organized, or one may want to monitor other sources not covered by the aforementioned products, or to keep control on flux thresholds on which follow-up alerts should be issued. In that spirit, it is worth noting that \citet{2011ICRC....8..137E} developed pipelines digesting \lat data for potential target of opportunity (ToO) observations with VERITAS. We~give another example below, which is in use in the \hess collaboration.

Within the \hess collaboration, the tool \texttt{FLaapLUC} (\textit{\underline{F}ermi}-\underline{L}AT \underline{a}utomatic \underline{a}perture \underline{p}hotometry \underline{L}ight \underline{C}$\leftrightarrow$\underline{U}rve) has been developed to identify flares in \lat data \citep{2018A+C....22....9L}. The purpose is to catch unusual activity in high-energy emitting sources as quickly as possible using \lat data, while~the flare is still on the go, in order to promptly organize ToO observations with \hess The method used here is also based on the aperture photometry\footnote{\href{https://fermi.gsfc.nasa.gov/ssc/data/analysis/scitools/aperture_photometry.html}{https://fermi.gsfc.nasa.gov/ssc/data/analysis/scitools/aperture\_photometry.html}}, which is computationally fast but inaccurate in terms of absolute flux scale, as mentioned above.

\begin{figure}[H]
\centering
\includegraphics[width=0.7\textwidth]{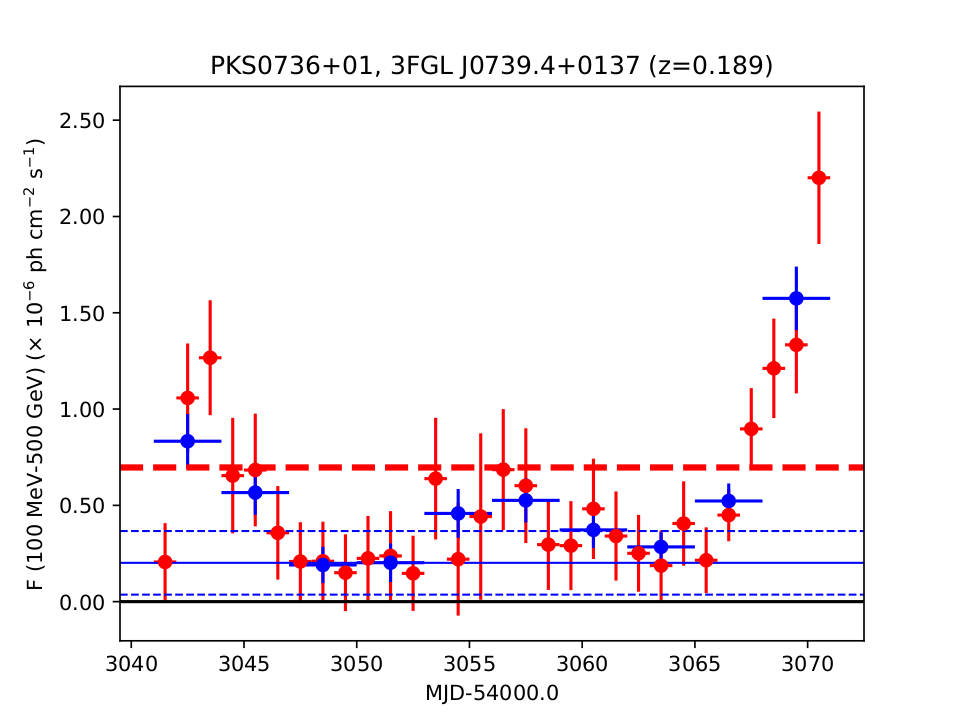}
\caption{One month segment of the \lat light curve from PKS\,0736$+$01 from \flaapluc. \hess observations were triggered and observations begun following the last measurement. The blue points show the data with bins of three days, while the red points represent the daily-binned data set. The~solid blue line shows the long-term (10 years) flux average with 1$\sigma$ fluctuations (dashed blue lines), and the dashed red line shows the flux alert threshold for this source.}
\label{fig-pks0736_flaapluc}
\end{figure}

More than 320 AGN are monitored daily with \flaapluc. This tool takes advantage of the whole mission data set of \lat to construct a long-term average flux ($\overline{F_\textnormal{LT}}$) for each source, which serves as the baseline for the identification of flares. Typically, if the last daily flux measurement of a source is above $\overline{F_\textnormal{LT}} + 3 \cdot \textnormal{RMS}(F_\textnormal{LT})$ (see Figure~\ref{fig-pks0736_flaapluc}), then an automatic alert is issued. In that case, a follow-up likelihood analysis is automatically processed to derive a precise spectral characterization. This alert is handled within the \hess AGN ToO group, which then evaluates the feasibility, and potentially triggers follow-up ToO observations with \hess \flaapluc veto alerts depending on the source distance and zenith angle at culmination, to take into account \g-ray absorption by the extragalactic background light \cite{2001ARA+A..39..249H}, as well as to account for source visibility at the \hess site. For more details on \flaapluc, whose code is available at \href{https://github.com/jlenain/flaapluc}{https://github.com/jlenain/flaapluc}, the reader is invited to \citet{2018A+C....22....9L}. \flaapluc has been in use in \hess since late 2012. Recently, \hess detected VHE flares following observations triggered with \flaapluc, for instance from 3C\,279 \citep{2018ATel.11239...1D}, TXS\,0506$+$056 along with a \lat alert \citep{2018ATel11419....1O}, PKS\,2022$-$07, PKS\,1749$+$096 or PKS\,0736$+$017 \citep{2017arXiv170800658C,2017arXiv170801083S}.

%%%%%%%%%%%%%%%%%%%%%%%%%%%%%%%%%%%%%%%%%%
\section{Conclusions}

Monitoring the high-energy sky is key for making new discoveries, either of previously unknown events, such as for the presumable coincidence of \g rays with neutrinos as observed in TXS\,0506$+$056~\cite{2018Sci...361.1378I}, or to follow up on flaring events in known AGN. To do so, all-sky or large field-of-view instruments, with high duty cycle, such as HAWC or \lat, are invaluable. Long-term monitoring programs as a strategy for observations, such as those performed by the FACT telescope \cite{2013JInst...8P6008A}, are also a key asset, since they provide regular observations of selected sources in an unbiased way {\cite{2017Galax...5...18T}}.

In the near future, the future CTA (Cherenkov Telescope Array) observatory is expected to provide, as parts of its key science projects, long-term monitoring of selected AGN, as well as a survey of a part of the extragalactic sky and of the Galactic plane (see \citep{2017arXiv170907997C}, for more details). Given its field of view of the order of a few degrees, CTA will presumably provide a shallow and wide survey of $\sim$25\% of the sky, with an expected sensitivity of more than ten times better than \lat or HAWC for steady sources in its core energy range (100\,GeV--10\,TeV). This~survey is expected to be completed within the first few years of operation of CTA, and thus complementary survey-capable facilities, such~as e-ASTROGAM \cite{2018JHEAp..19....1D} or AMEGO\footnote{See \href{https://asd.gsfc.nasa.gov/amego/}{https://asd.gsfc.nasa.gov/amego}}, are necessary to take such a survey over in the next couple of decades. Such satellite projects being still in the proposal stage, assuring that \textit{Fermi} keeps flying in the next years is essential for the community.

%%%%%%%%%%%%%%%%%%%%%%%%%%%%%%%%%%%%%%%%%%
% \section{Patents}
% This section is not mandatory, but may be added if there are patents resulting from the work reported in this manuscript.

%%%%%%%%%%%%%%%%%%%%%%%%%%%%%%%%%%%%%%%%%%
\vspace{6pt}

%%%%%%%%%%%%%%%%%%%%%%%%%%%%%%%%%%%%%%%%%%
%% optional
%\supplementary{The following are available online at \linksupplementary{s1}, Figure S1: title, Table S1: title, Video S1: title.}

% \supplementary{The following are available at \linksupplementary, Figure S1: title, Table S1: title, Video S1: title. A supporting video article is available at doi: link.}

%%%%%%%%%%%%%%%%%%%%%%%%%%%%%%%%%%%%%%%%%%
% \authorcontributions{For research articles with several authors, a short paragraph specifying their individual contributions must be provided. The following statements should be used “conceptualization, X.X. and Y.Y.; methodology, X.X.; software, X.X.; validation, X.X., Y.Y. and Z.Z.; formal analysis, X.X.; investigation, X.X.; resources, X.X.; data curation, X.X.; writing—original draft preparation, X.X.; writing—review and editing, X.X.; visualization, X.X.; supervision, X.X.; project administration, X.X.; funding acquisition, Y.Y.”, please turn to the  \href{http://img.mdpi.org/data/contributor-role-instruction.pdf}{CRediT taxonomy} for the term explanation. Authorship must be limited to those who have contributed substantially to the work reported.}

%%%%%%%%%%%%%%%%%%%%%%%%%%%%%%%%%%%%%%%%%%
\funding{This research received no external funding.}
% \funding{Please add: ``This research received no external funding'' or ``This research was funded by NAME OF FUNDER grant number XXX.'' and  and ``The APC was funded by XXX''. Check carefully that the details given are accurate and use the standard spelling of funding agency names at \url{https://search.crossref.org/funding}, any errors may affect your future funding.}

%%%%%%%%%%%%%%%%%%%%%%%%%%%%%%%%%%%%%%%%%%
\acknowledgments{The support of the Namibian authorities and of the University of Namibia in facilitating the construction and operation of \hess is gratefully acknowledged, as is the support by the German Ministry for Education and Research (BMBF), the Max Planck Society, the German Research Foundation (DFG), the Helmholtz Association, the Alexander von Humboldt Foundation, the French Ministry of Higher Education, Research and Innovation, the Centre National de la Recherche Scientifique (CNRS/IN2P3 and CNRS/INSU), the Commissariat à l'énergie atomique et aux énergies alternatives (CEA), the U.K. Science and Technology Facilities Council (STFC), the Knut and Alice Wallenberg Foundation, the National Science Centre, Poland Grant No. 2016/22/M/ST9/00382, the South African Department of Science and Technology and National Research Foundation, the University of Namibia, the National Commission on Research, Science \& Technology of Namibia (NCRST), the Austrian Federal Ministry of Education, Science and Research and the Austrian Science Fund (FWF), the Australian Research Council (ARC), the Japan Society for the Promotion of Science and by the University of Amsterdam. We appreciate the excellent work of the technical support staff in Berlin, Zeuthen, Heidelberg, Palaiseau, Paris, Saclay, Tübingen and in Namibia in the construction and operation of the equipment. This work benefitted from services provided by the \hess Virtual Organisation, supported by the national resource providers of the EGI Federation.

This research has made use of NASA's Astrophysics Data System, of the SIMBAD database, operated at CDS, Strasbourg, France, and of the TeVCat
online source catalog (\href{http://tevcat.uchicago.edu}{http://tevcat.uchicago.edu}). Part of this work is based on archival data, software or online services provided by the Space Science Data Center---ASI.
}

%%%%%%%%%%%%%%%%%%%%%%%%%%%%%%%%%%%%%%%%%%
\conflictsofinterest{The author declares no conflict of interest.
% Declare conflicts of interest or state ``The authors declare no conflict of interest.'' Authors must identify and declare any personal circumstances or interest that may be perceived as inappropriately influencing the representation or interpretation of reported research results. Any role of the funding sponsors in the design of the study; in the collection, analyses or interpretation of data; in the writing of the manuscript, or in the decision to publish the results must be declared in this section. If there is no role, please state ``The founding sponsors had no role in the design of the study; in the collection, analyses, or interpretation of data; in the writing of the manuscript, or in the decision to publish the results''.
}

\externalbibliography{yes}
\bibliography{review_monitoring_jpl.bbl}

%%%%%%%%%%%%%%%%%%%%%%%%%%%%%%%%%%%%%%%%%%
%% optional
% \sampleavailability{Samples of the compounds ...... are available from the authors.}

%% for journal Sci
%\reviewreports{\\
%Reviewer 1 comments and authors’ response\\
%Reviewer 2 comments and authors’ response\\
%Reviewer 3 comments and authors’ response
%}

%%%%%%%%%%%%%%%%%%%%%%%%%%%%%%%%%%%%%%%%%%
\end{document}